\documentclass[nonacm,sigconf]{acmart}

\AtBeginDocument{%
  }

\usepackage{color}
\setcopyright{acmlicensed}
\copyrightyear{2024}
\acmYear{2024}
\acmDOI{XXXXXXX.XXXXXXX}

\usepackage{cite}
\usepackage{xcolor}
\usepackage{enumitem}
\usepackage{amsmath,amsfonts}
\usepackage{algorithmic}
\usepackage{graphicx}
\usepackage{textcomp}
\usepackage[most]{tcolorbox}
\usepackage[framemethod=TikZ]{mdframed}
\usepackage{natbib}
\setcitestyle{numbers,sort&compress}

\mdfdefinestyle{exampledefault}{
    skipabove=0.2cm,  
    middlelinecolor = black,
    middlelinewidth=0.5pt,
    linecolor=black,
    backgroundcolor=gray!5,
    roundcorner = 5pt,
}
\begin{document}

\title{CoDefeater: Using LLMs To Find Defeaters in Assurance Cases}

\author{Usman Gohar}
\affiliation{%
 \institution{Iowa State University}
 \city{Ames}
 \state{Iowa}
 \country{USA}}
 \email{ugohar@iastate.edu}

\author{Michael C. Hunter }
\affiliation{%
 \institution{Iowa State University}
  \streetaddress{30 Shuangqing Rd}
 \city{Ames}
 \state{Iowa}
 \country{USA}}
 \email{mchunter@iastate.edu}

\author{Robyn R. Lutz}
\affiliation{%
 \institution{Iowa State University}
  \streetaddress{1 Th{\o}rv{\"a}ld Circle}
 \city{Ames}
 \state{Iowa}
 \country{USA}}
\email{rlutz@iastate.edu}

\author{Myra B. Cohen}
\affiliation{%
 \institution{Iowa State University}
 \city{Ames}
 \state{Iowa}
 \country{USA}}
\email{mcohen@iastate.edu}

\begin{abstract}
 
Constructing assurance cases is a widely used, and sometimes required, process toward demonstrating that safety-critical systems will operate safely in their planned environment. To mitigate the risk of errors and missing edge cases, the concept of defeaters - arguments or evidence that challenge claims in an assurance case - has been introduced. Defeaters can provide timely detection of weaknesses in the arguments, prompting further investigation and timely mitigations. 
However, capturing defeaters relies on expert judgment, experience, and creativity and must be done iteratively due to evolving requirements and regulations. This new ideas paper proposes CoDefeater, an automated process to leverage large language models (LLMs) for finding defeaters. Initial results on two systems show that LLMs can efficiently find known and unforeseen feasible defeaters to support safety analysts in enhancing the completeness and confidence of assurance cases.

\end{abstract}



\maketitle

\section{Introduction}

Safety-critical systems have become deeply integrated into many societal domains, including healthcare, transportation, energy, and aviation \citep{10.1145/3342481,dey2018medical,gunes2014survey}. Failures in these systems can lead to catastrophic consequences for human safety, including fatalities and environmental and property damage \citep{rushby1994critical}. This has led to an increased focus on their dependence, reliability, and safety \citep{rausand2014reliability,knight2002safety}. Many systems must comply with regulations \citep{parliament2016general,breaux2006towards}, provide evidence of safety, and undergo rigorous certification processes \citep{johnson1998178b,palin2011iso} for approval from regulatory bodies. \emph{Assurance cases} (ACs) have emerged as a common practice for this purpose, facilitating the verification of system correctness and the validation of specific claims regarding safety, security, and trustworthiness, among others \citep{maksimov2018two,bloomfield2020assurance, ACWG21}.

\begin{figure}[t]
    \centering
    \includegraphics[scale = 0.85]{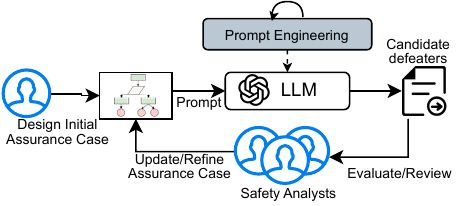}
    \caption{Overview of CoDefeater.}
    \label{fig:system}
\end{figure}

An assurance case is a structured hierarchy of claims and arguments supported by evidence that a system will function as intended in a specified environment \citep{rushby2015interpretation,bloomfield2020assurance}. Several formal notations (e.g., Goal Structuring Notation (GSN) \citep{kelly2004goal}, Claims-Arguments-Evidence (CAE) \citep{Adelard}, and Eliminative Argumentation (EA) \citep{goodenough2015eliminative}), 
along with tools \citep{denney2012advocate,10311213,maksimov2018two}, have been proposed. However, concerns arise over their completeness, uncertainty, and soundness for cyber-physical systems
\citep{Greenwell2004ATO,duan2017reasoning}, leading to false confidence and catastrophic failures \citep{8590172}. For example, failure of the minimum safe altitude warning system that led to a major aviation accident was attributed to incomplete and flawed reasoning in the safety case \citep{greenwell2004failure}. 

To enhance the robustness of assurance cases, it is critical to identify and mitigate their \emph{defeaters} (also known as assurance weakeners). Defeaters highlight gaps in evidence or reasoning that undermine the validity of claims in the assurance case \citep{hawkins2011new}.
An example of a defeater, drawn from the assurance case for the safe operation of an sUAS (small Uncrewed Aircraft System) battery, challenges the assurance case's claim that ``The sUAS has enough battery charge to complete its mission." The defeater casting doubt on this claim is, ``Unless the battery monitor is not calibrated/inaccurate." The defeater serves to record, within the assurance case itself, the analyst's challenge to the validity of the claim. 

Various approaches have been proposed to identify and mitigate defeaters  \citep{graydon2017investigation,khakzad}. However, manually creating defeaters is a labor-intensive and time-consuming process \citep{nair2014extended,DBLP:conf/icse/MenghiVSRJC23}, relying heavily on safety analysts' judgment, experience, creativity, and understanding of the system. This can lead to confirmation bias \citep{10311213, bloomfield2021safety}. As assurance cases evolve with new standards and technological advances, ongoing efforts are focused on formal and semi-automated approaches for detecting and managing defeaters \citep{diemert2023incremental, 10.1145/2976767.2976792}. Best practices build assurance cases incrementally, so automating all/or part of this process is important \citep{warg2019continuous, diemert2023incremental}. 

Large Language Models (LLMs) are increasingly automating software engineering tasks like test generation and defect detection \citep{Hou2023LargeLM,ahmed2022few}. 
In particular, these models have become valuable in tasks requiring complex understanding, such as vulnerability detection, requirements elicitation, and code generation \citep{zhang2023prompt,chen2021evaluating,arora2024advancing}. Moreover, LLMs have shown promise in automating evaluation tasks and acting as surrogate evaluators \citep{chiang2023can,zheng2024judging}. Consequently, we explore whether LLMs' capabilities can be harnessed to automate defeater analysis toward the completeness, soundness, and confidence of assurance cases. Despite calls for more research into LLMs' ability to identify defeaters \citep{shahandashti2024evaluating,Viger23}, no study has been conducted to evaluate and investigate their effectiveness.

This new ideas paper presents the first empirical investigation of the feasibility and utility of LLMs for identifying defeaters, using a process we call \emph{CoDefeater}. It evaluates their potential to aid practitioners and safety analysts in the iterative human-in-the-loop process of finding defeaters, as shown in Figure \ref{fig:system}. We evaluate the performance of an LLM (ChatGPT) in automated defeater analysis on two complex real-world case studies. Our experimental results suggest that CoDefeater is a promising approach for identifying and generating novel assurance case defeaters. Overall, this work makes three key contributions. \textbf{1)} To the best of our knowledge, we provide the first empirical results from an investigation of the effectiveness and usefulness of an LLM (GPT 3.5) in identifying and creating defeaters for real-world assurance cases. \textbf{2)} We provide a new assurance case fragment with defeaters that can be leveraged for further research on automated defeater identification techniques. \textbf{3)}. Based on our findings, we outline current challenges and directions for future work. All experimental artifacts are available here: https://gitlab.com/anonymousdot/codefeater.
\section{Background and Related Work}
\label{sec:background}

\begin{figure}[]
    \centering
    \includegraphics[scale = 0.70]{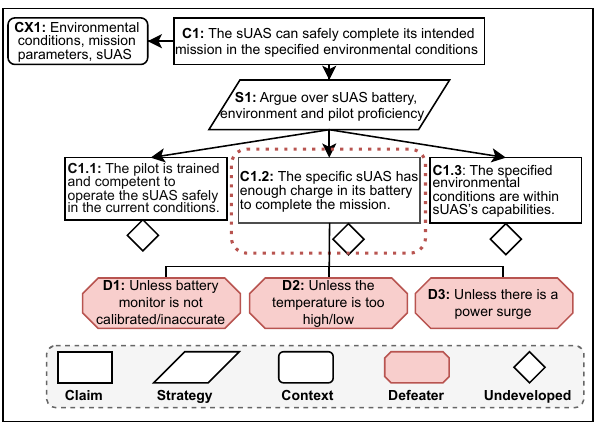}
    \caption{An Assurance Case fragment with three example defeaters for Claim 1.2.}
    \label{fig:AC}
    \vspace{-3.5mm}
\end{figure}

Assurance case arguments typically adopt an inductive approach, where sub-claims offer direct evidence to support the parent claim but do not ensure it with certainty \citep{bloomfield2020assurance,chechik2019uncertain}. Focusing solely on proving a claim may introduce confirmation bias, as exemplified by the Nimrod aircraft crash \citep{cave2006independent}. Recent approaches have therefore embraced defeasible reasoning, which acknowledges that arguments about system properties in practice are inherently defeasible~\citep{Goodenough2012TowardAT,koons2005defeasible,pollock1987defeasible}. \textit{Defeaters are potential doubts or objections that challenge the validity of a claim, reflecting gaps in evidence and reasoning}~\citep{Goodenough2012TowardAT, Koopman22,duan2017reasoning, Murugesan23}. 
Figure \ref{fig:AC} shows a fragment of an assurance case for an sUAS battery, with examples of defeaters (red boxes).

Defeaters are typically represented using EA notation \citep{goodenough2015eliminative} but have also been integrated into GSN and CAE notations \citep{hawkins2011new,bloomfield2020assurance}. Hence, this study does not aim to evaluate LLMs' performance in identifying defeaters in any specific notation or semantic accuracy but rather to investigate the feasibility of this approach. Finally, the indefeasibility criterion requires a thorough search for defeaters in an assurance case \citep{bloomfield2020assurance}, motivating our investigation into LLMs' potential to assist practitioners and safety analysts in identifying and generating novel defeaters. 

In software engineering, LLMs are increasingly employed to assist developers and automate tasks such as discovering requirements, code generation, testing, and program synthesis \citep{Hou2023LargeLM, arora2024advancing}. \citet{diemert2023can} have reported ChatGPT's effectiveness in hazard analysis for safety-critical systems, highlighting their potential to assist human analysts. In the context of software assurance cases, \citet{sivakumar2023gpt} assessed LLM's (GPT-4) proficiency in understanding GSN representations and its performance in constructing safety cases. \citet{Viger23} proposed using LLMs to identify defeaters; however, they did not report an empirical evaluation of their capabilities. \citet{shahandashti2024evaluating} explored LLM's (GPT-4) understanding of EA notation and defeater concepts, with empirical validation left as future work. We aim to address this gap by examining LLM performance in identifying defeaters for assurance cases in two case studies.

The Machine Learning (ML) community also has explored the reasoning abilities of LLMs \citep{Zheng2023JudgingLW,dai2023llm,Wang2024ExploringTR}. This line of inquiry investigates how LLMs fare in open-ended tasks and their effectiveness in assisting humans.
Findings have indicated that LLMs can demonstrate consistent responses exhibiting similarities with human evaluations, suggesting their potential as automated tools \citep{chiang2023can}.

\section{Methodology}

We conducted a preliminary exploratory study toward answering the following two research questions:

\begin{itemize}[leftmargin=*]
    \item \textbf{RQ1: (Effectiveness).} 
    \textit{How effective are LLMs in identifying and analyzing defeaters in assurance cases?} 
    \item \textbf{RQ2: (Utility).}  
    \textit{Can LLMs support practitioners in generating novel and meaningful defeaters?}
\end{itemize}

\subsection{Experimental Setup}

\noindent \textbf{Datasets.} We performed our experiments on two assurance cases. The first is for the \textit{CERN Large Hadron Collider (LHC) Machine Protection System (MPS)}, which provides assurance that the MPS will prevent damage to the LHC from unstable, high-energy particle beams \citep{Rees_Lippelt, millet2023assurance}. The assurance case for the CERN LHC uses EA notation, and we extracted claim nodes with their corresponding defeaters for our experiments. 
The second assurance case is a fragment of a larger one that our team has recently created for small Uncrewed Aircraft Systems (sUAS)\footnote{Anonymized for submission} using GSN. It addresses the claim ``\textit{the sUAS has enough charge in its battery to complete the mission}" (see claim 1.2 in Figure \ref{fig:AC}). We include both the assurance case fragment and corresponding defeaters in the supplementary material\footnote{https://gitlab.com/anonymousdot/codefeater}. The real-world complexity of these systems and the availability of defeaters made them suitable for our preliminary experiments. Table \ref{table:Assurance case stats} provides a summary of the assurance cases. 

\noindent \textbf{Model.} We used  ChatGPT (GPT-3.5) for our study, specifically model \emph{GPT-3.5-turbo} released by Open AI \citep{brown2020language}.

\begin{table}[]
\caption{Number of claims and defeaters in the ACs.}
\label{table:Assurance case stats}
\begin{tabular}{ccc}
\toprule
         \textbf{Node Type} & \textbf{LHC MPS} & \textbf{sUAS Battery} \\ \midrule
Claims    & 61                   & 6                                                 \\
Defeaters & 103                   &  20  \\ \bottomrule                                        
\end{tabular}
\end{table}

\subsection{Prompt Design}
\begin{figure}[]
    \centering
    \includegraphics[width = \columnwidth]{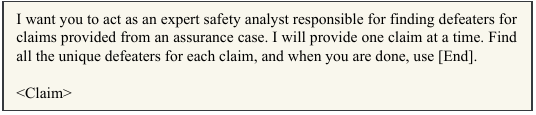}
    \caption{The system prompt used in the study.}
    \vspace{-2mm}
    \label{fig:prompt}
\end{figure}

An effective prompt design is crucial for achieving good performance, as the choice of prompts significantly impacts the quality, relevance, and accuracy of the LLM's response \citep{chen2023unleashing}. It involves crafting a system prompt to establish the context and prepare the LLM for the task, along with a user prompt that contains the specific task request \citep{openAI}. 
We designed the system prompt following OpenAI's best practices \citep{openAI} and relevant literature in software engineering \citep{zhou2024large, chen2023use} and open-ended evaluation tasks \citep{chiang2023can}.  Our process identified role-based system prompts \citep{kong2023better} as the most effective approach. Figure \ref{fig:prompt} shows the system prompt used in our study. 

Several user prompting techniques have been proposed, e.g., zero-shot, one-shot, few-shot, and chain-of-thought \citep{chen2023unleashing}. Zero-shot learning involves providing the model with only the task description (system prompt), without examples of unseen tasks to learn from. In contrast, one-shot and few-shot learning conditions the model on one or more examples in the prompt, respectively \citep{brown2020language}. 

For our preliminary study, we adopted the zero-shot learning setting. This approach both (1) facilitates the immediate, off-the-shelf application of LLMs, eliminating the need for computationally expensive fine-tuning procedures, and (2) is naturally suited to scenarios such as ours, with limited data availability for training or fine-tuning \citep{kojima2022large}. Each prompt was presented independently to the model to avoid influencing subsequent responses, allowing us to assess its standalone capabilities \citep{Chen2023HowIC}.

\subsection{Evaluation Criteria}
\label{sec:eval}

\begin{figure}[t]
    \centering
    \includegraphics[width = \columnwidth]{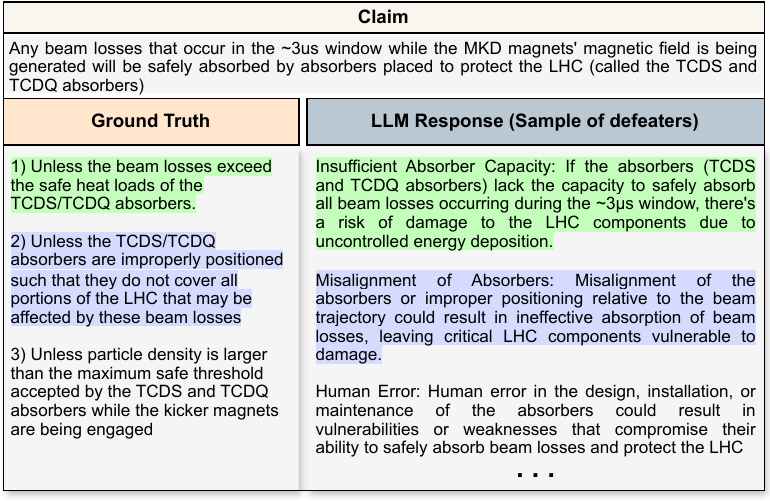}
    \caption{A sample claim from the LHC assurance case, together with the claim's ground-truth defeaters (left) and LLM-generated (ChatGPT) defeaters (right), color-coded to represent the level of agreement between the two: complete match (green), partial match (blue), and no match (no color). }
    \vspace{-6mm}
    \label{fig:sample}
\end{figure}

Due to the complexity and open-ended nature of the task (e.g., varying response length vs. ground truth, and subjectivity), automatic evaluation metrics were not suitable \citep{chiang2023can}. Therefore, we relied on human evaluation for assessing LLM performance. For RQ1, we used a deductive coding approach \citep{kulatunga2007structuring,diemert2023can}, where the first two authors independently reviewed responses and categorized them as \textit{complete match}, \textit{partial match}, or \textit{no match} based on similarity to ground-truth defeaters. We used the defeaters in the LHC assurance case as the ground truth.  For the battery assurance case, a set of ground-truth defeaters was provided by one of the authors (independently)  familiar with the domain, following best practices \citep{millet2023assurance}. 
For RQ2, the responses were evaluated for being \textit{reasonable} \citep{chen2023use}, i.e., the defeater could reasonably be in the ground truth but had been overlooked. This aimed to assess the LLM's capability to identify novel defeaters. Figure \ref{fig:sample} shows an example of each type of match. Next, the reviewers met to discuss and finalize their assigned codes. In the case of post-discussion disagreement, if one reviewer labeled a response as a partial match and the other as a complete match, we categorized it as a partial match to avoid confirmation bias \citep{gemalmaz2021accounting}. In the one instance where one reviewer indicated no match while the other identified a partial or complete match, it was discarded. Last, we calculated inter-rater agreement using Cohen's Kappa \citep{cohen1960coefficient} to evaluate the consistency and reliability of the coding process.

\subsection{Threats to Validity}
There is potential subjectivity in the qualitative evaluation of LLM performance on defeater identification. To address that, two authors independently coded the LLM responses, following best practices \citep{braun2006using,gohar2024towards}, and held multiple discussions to avoid misinterpretations.  We also computed Cohen's kappa \citep{cohen1960coefficient}, which indicated substantial inter-reviewer agreement. To avoid confirmation bias, disagreements were coded as partial or no match. The non-deterministic nature of LLMs and different versions might produce slightly different responses; however, we used a single version of ChatGPT. Finally, the preliminary results presented here lead us to propose that the use of LLMs to generate defeaters merits further work; however, generalizability awaits larger studies with improved LLMs. 

\newcounter{finding}
\section{Results}

In this section, we present the key findings of our experiments, grouped by our research questions. 

\subsection{(RQ1): Effectiveness in Identifying Defeaters}
\stepcounter{finding}
\begin{mdframed}[style = exampledefault]
\noindent \textbf{Finding \thefinding:} 
The LLM displayed promising zero-shot capabilities for defeater analysis in assurance cases.
\end{mdframed}

\begin{figure}[t]
    \centering
    \includegraphics[scale = 0.35]{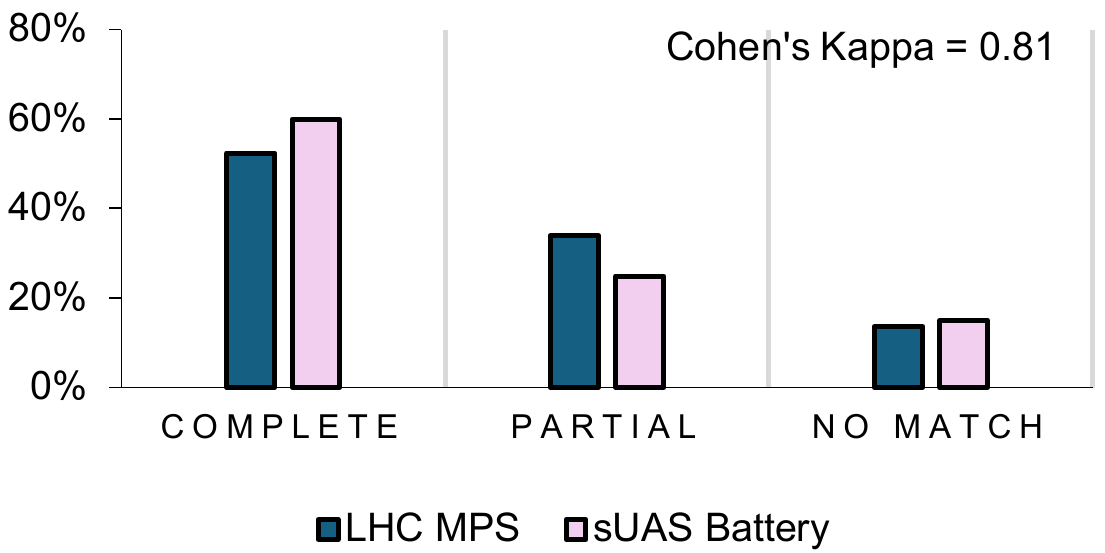}
    \caption{(Performance). Distribution of defeaters across coding categories. Cohen's kappa showed almost perfect agreement beyond chance.}
    \vspace{-2mm}
    \label{fig:result1}
\end{figure}

Figure \ref{fig:result1} presents the results of our closed coding process, which assessed the GPT-3.5 model's performance in identifying defeaters for the two real-world assurance cases in a zero-shot setting. Our experiments showed that the model demonstrated unexpectedly good zero-shot capabilities for defeater analysis. Specifically, it completely identified more than half of all defeaters and partially identified more than a third in both datasets. Fewer than $15\%$ of the defeaters created by human analysts were not identified at all by the LLM. These results were noteworthy given the complexity of the systems under study and the lack of system information, domain knowledge, or few-shot examples provided to the LLM. 
Additionally, the total coded responses achieved high inter-rater agreement, with a Cohen's Kappa score of 0.81, 
\citep{cohen1960coefficient, mchugh2012interrater}. Finally, all of these defeaters were identified in the first prompt, illustrating the convenience of the approach. Among the unidentified defeaters, some required specific domain or system knowledge not provided to the model, indicating areas for potential improvement. 

\stepcounter{finding}
\begin{mdframed}[style = exampledefault]
\noindent \textbf{Finding \thefinding:} 
The LLM struggled with defeaters that challenged implicit assumptions.
\end{mdframed}

We conducted a manual analysis of the unidentified defeaters (n = 17) to investigate whether there were any patterns behind the LLM's failure. Interestingly, we found that the model struggled to identify those defeaters that implicitly challenged the truth of an assumption. For example, for the claim "\textit{The BICs will not produce a FALSE BEAM-PERMIT to trigger a beam dump, unless a loss of the high-frequency signal (10Mhz) in either Beam Permit Loop (A and B) is detected..}.", a (ground-truth) defeater in the LHC assurance case questions the assumption that the 10Mhz signal is the right signal to monitor. In other words, the presence of another similar high-frequency signal might lead to a false indication of TRUE BEAM-PERMIT. Unlike the analysts, the LLM did not question the underlying assumption in the claim and thus did not identify the defeater. Consequently, future work should explore integrating external knowledge sources since, in similar tasks, it significantly enhances LLM's performance \citep{zhou2024large,peng2023check}. 

\subsection{(RQ2): Utility in Generating Novel Defeaters}

To answer RQ2, we evaluated the LLM's performance on the sUAS battery's assurance case, where we have the necessary domain knowledge. Using the same prompting method, we iteratively requested additional defeaters to assess its capacity to generate novel defeaters beyond the ones that had been identified in the building of the assurance case.

\stepcounter{finding}
\begin{mdframed}[style = exampledefault]
\noindent \textbf{Finding \thefinding:} LLMs can support practitioners in providing useful and novel defeaters.
\end{mdframed}

The LLM output was a useful source of five novel defeaters, each of which was feasible upon further investigation using our evaluation criteria. These were: (1) an unexpected power drain due to an onboard component failure; (2) an emergency external to the sUAS that forced the sUAS into a longer flight; (3) a missed waypoint to which the pilot had to return; (4) unexpected power drain arising from ongoing efforts to recover a lost GPS; and (5) external interference that the sUAS had to dodge repeatedly. The last one was interesting to us because the LLM  gave as an example of interference that birds might attack the sUAS.  This, in fact, happens quite often and is dangerous \citep{Birds}. The response shows how an LLM can offer a creative perspective that catches missing edge cases.  

\section{Discussion}

Based on our findings, we highlight several challenges and opportunities for an LLM-based process to help find defeaters. 

\noindent \textbf{Designing better prompts.} Prompt designing has been shown to significantly impact LLM performance \citep{kong2023better,chen2023unleashing}. Many prompting methods have been proposed, and further investigation for suitability to defeater analysis is needed. Moreover, our study revealed that the LLM can generate creative, redundant, and far-fetched scenarios (e.g., defeaters due to budget constraints). Balancing LLM creativity with defeater relevance poses an important challenge. 

\noindent \textbf{Rationale behind defeaters. } In our experiments, we found that the LLM responses not only identified defeaters but also provided helpful rationale and examples. For instance, if the ground truth defeater stated, "\textit{Unless there are incorrect readings}," the LLM suggested, "\textit{The sensors may not be properly calibrated, leading to inaccurate readings}." These explanations can assist analysts in understanding and analyzing both a defeater's feasibility and its potential mitigations. Investigating explainable prompting techniques such as Chain-Of-Thought \citep{chen2023unleashing} is an important next step.

\noindent \textbf{Towards incremental assurance using LLMs.} Our study focused on single claims and associated defeaters.
Future research should evaluate the performance of LLMs on a combination of claims. It will be interesting to investigate whether LLMs can identify the impact of defeaters on multiple claims and assess if the provided evidence adequately addresses them. This direction will require developing detailed data for evidence analysis and exploring prompts specifically designed for this purpose. 
\section{Conclusion and Future Work}

We have presented CoDefeater, a process for automated defeater discovery in assurance cases using LLMs (GPT-3.5). Our evaluation on two real-world case studies demonstrated the LLM's zero-shot capabilities in identifying defeaters and its potential to support practitioners in an iterative human-in-the-loop process. We make available the portion of a new assurance case and its ground-truth defeaters used in our experiments for other researchers. Future work will expand beyond the zero-shot setting to explore one-shot and few-shot learning approaches for improved performance. Additionally, fine-tuning LLMs on assurance cases presents an avenue to improve their performance. Our study provides preliminary results as a starting point for future research to explore the role of LLMs as a tool to assist with the identification of defeaters toward the development of improved assurance cases.

\begin{acks}
This work was funded by grant 80NSSC23M0058 from the National Aeronautics and Space Administration 
(NASA). 
\end{acks}

\bibliographystyle{ACM-Reference-Format}
\bibliography{sample-base}

\end{document}